\documentstyle[12pt,preprint,epsf]{aastex}

\def\agt{\mathrel{\raise.3ex\hbox{$>$}\mkern-14mu\lower0.6ex\hbox{$\sim$}}}
\def\alt{\mathrel{\raise.3ex\hbox{$<$}\mkern-14mu\lower0.6ex\hbox{$\sim$}}}

\newcommand{\beq}{\begin{equation}}
\newcommand{\eeq}{\end{equation}}
\newcommand{\beqn}{\begin{eqnarray}}
\newcommand{\eeqn}{\end{eqnarray}}

\begin{document}

\title{Collapse of a Rotating Supermassive Star to a
Supermassive Black Hole: Fully Relativistic Simulations}

\author{Masaru Shibata \altaffilmark{1}
and Stuart L. Shapiro \altaffilmark{2,3}}

\affil
{\altaffilmark{1} 
Graduate School of Arts and Sciences, 
University of Tokyo,
\break
Komaba, Meguro, Tokyo 153-8902, Japan}

\affil
{\altaffilmark{2} 
Department of Physics, University of Illinois at Urbana-Champaign,
\break
Urbana, IL 61801-3080}
\affil
{\altaffilmark{3} 
Department of Astronomy and NCSA, University of Illinois at
Urbana-Champaign,
\break
Urbana, IL 61801-3080}



\begin{abstract}
We follow the collapse in axisymmetry of a uniformly rotating,
supermassive star (SMS) to a supermassive black hole (SMBH) 
in full general relativity. The initial SMS of 
arbitrary mass $M$ is marginally unstable to radial collapse 
and rotates at the mass-shedding limit. The collapse proceeds 
homologously early on and results in the appearance of an apparent 
horizon at the center. Although our integration terminates before 
final equilibrium is achieved, we determine that the final black hole 
will contain about $90\%$ of the total mass of the system and have a 
spin parameter $J/M^2 \sim 0.75$. The remaining gas forms a 
rotating disk about the nascent hole.
\end{abstract}




\section{Introduction}

Recent observations provide increasingly
strong evidence that supermassive black
holes (SMBHs) of mass $\sim 10^6 - 10^{10}M_{\odot}$ exist and that
they are the central engines that power active galactic
nuclei (AGNs) and quasars (Rees 1998, 2001). 
The dynamical formation of SMBHs, as well as the inspiral, collision and
merger of binary SMBHs, are promising sources of long-wavelength 
gravitational waves for the proposed Laser Interferometer Space Antenna (LISA) 
(Thorne 1995; Schutz 2001). However, the actual
scenario(s) by which SMBHs form is(are) still uncertain. Viable stellar 
dynamical and hydrodynamical routes leading to the formation of SMBHs 
have been proposed (e.g., Begelman \& Rees 1978, Rees 1984,
Shapiro \& Teukolsky 1985, Quinlan \& Shapiro 1990, Rees 1998, 2001;
Balberg \& Shapiro 2002). In typical hydrodynamical scenarios, 
a supermassive gas cloud is build up from the multiple collisions of 
stars or small gas clouds in stellar clusters to form a supermassive star (SMS)
(Begelman \& Rees 1978).  A supermassive gas cloud might
be formed from scattered gas in the bulge that falls into the
central region due to radiation drag (Umemura 2001). SMSs 
ultimately collapse to black holes following quasi-stationary 
cooling and contraction to the onset of radial instability 
(Zel'dovich \& Novikov 1971; Shapiro \& Teukolsky 1983). This scenario 
for forming a SMBH following SMS collapse was investigated in the 
1960s and 1970s, but the studies
treated nonrotating configurations and assumed spherical symmetry, or
employed approximate analytical models (see, e.g., 
Baumgarte \& Shapiro 1999 for a review and references).

However, SMSs are likely to be rapidly rotating
(see, e.g., Loeb \& Rasio 1994).
Baumgarte \& Shapiro (1999) recently performed a detailed
numerical analysis of the structure and stability of a rapidly rotating
SMS in equilibrium. Assuming the viscous or magnetic 
braking timescale for angular momentum transfer 
is shorter than the evolution timescale
of a typical SMS (Zel'dovich \& Novikov 1971; see New and Shapiro 2001 for
an alternative), the star will settle into rigid rotation 
and evolve to the mass-shedding limit following cooling and contraction. 
(At mass-shedding, the angular velocity of gas at the equator equals the 
Kepler velocity). They found that all stars at
the onset of quasi-radial collapse have an  
equatorial radius $R \approx 640GM/c^2$ and a nondimensional
spin parameter $q \equiv c J/GM^2 
\approx 0.97$. Here $J$, $M$, $c$, and $G$ 
are spin, gravitational mass, light velocity and gravitational constant. 
(Hereafter we adopt gravitational units and set $c=G=1$).
Because of the large value of $q$, it is uncertain whether the rotating SMS 
collapses directly to a black hole or forms a disk.
Saijo et al. (2002) investigated the collapse of a rotating SMS 
in a post-Newtonian (PN) approximation, and concluded that 
effects of rotation do not halt the collapse and that 
a SMBH is likely to be formed. However, a PN calculation
cannot follow collapse into the strong-field regime and
cannot rigorously address the possibility of black hole 
formation and growth, and the final outcome. 

To clarify whether a SMBH 
forms as the endpoint of rapidly rotating SMS collapse and, if it does, 
to determine the final hole parameters, we performed a fully general 
relativistic numerical simulation of the collapse in axisymmetry.
We demonstrate that the collapse indeed leads directly to a SMBH
of moderately rapid rotation, with $q \sim 0.75$.

\section{Computational Set-Up}

Simulations were performed using an axisymmetric 
code in full general relativity (Shibata 2000).
This code was constructed 
from a 3D code (Shibata 1999b) using
the so-called ``cartoon method'' in numerical
relativity (Alcubierre et al. 2001): We solve the Einstein
field equations in Cartesian coordinates with
a grid of size $(N, 3, N)$ in $(x, y, z)$,
covering a computational domain 
$0\leq x, z \leq L$ and $-\Delta y \leq y \leq \Delta y$. 
Here $N$ is a constant ($\gg 1$), $L$ the location of
the outer boundary and $\Delta y$ is the grid spacing in
the $y$-direction, with an an axisymmetric boundary condition imposed 
at $y=\pm \Delta y$; the spin axis is along $z$.
We solve the hydrodynamic equations with an axisymmetric code in
cylindrical coordinates. 
This hydrodynamical code
was calibrated by numerous test calculations, including 
spherical collapse of dust, the stability of spherical stars,
mode analysis of spherical stars, 
and the evolution of rotating stars (Shibata 1999b). 
We adopt maximal time slicing and approximate
minimal distortion as coordinate gauge conditions throughout the simulation 
(Shibata 1999a, b). 
Formation of a black hole is determined by finding an apparent horizon.

Violations of the Hamiltonian constraint and conservation of mass and 
angular momentum are monitored as numerical accuracy check during 
the simulation. Total angular momentum $J$ and 
total baryon rest-mass $M_*$ should be 
strictly conserved in axisymmetry. The gravitational mass $M$ is not
conserved, due to the emission of
gravitational radiation, but the decrease in $M$ is very small.

Since the SMS spins up during the collapse, a
nonaxisymmetric instability could arise. However, the PN study by
Saijo et al. (2002) indicates that such an instability is not
excited, at least when the equatorial radius of
the collapsing star exceeds $10M$. 
By restricting our analysis to axisymmetry,
we can improve our resolution of the strong-field, central region where the
black hole forms by a factor of $\agt 10$. 

To model the initial equilibrium SMS, we adopt a polytropic equation of state
with the adiabatic index $\Gamma=4/3$, setting $P = K\rho^{4/3}$, 
where $P$ and $\rho$ are the pressure and rest-mass density. This 
prescription is appropriate for a radiation-dominated SMS equation of state.
Here $K$ is a constant whose value determines the mass; we 
scale out the mass by setting $K=1$ (Baumgarte \& Shapiro 1999).

Although a spherical SMS with 
the adopted equation of state is unconditionally unstable
to collapse, rotation can stabilize the star. 
We focus only on a rigidly rotating SMS at the mass-shedding limit. 
According to Baumgarte \& Shapiro (1999), 
rotating SMSs with equatorial radii $R < 640M$ are unstable against collapse. 
At $R \approx 640M$, 
the ratio of the rotational kinetic energy to the gravitational binding 
energy $T/W$ is about $0.009$ and $q \approx 0.97$.

We chose a SMS with $R \approx 620M$, 
a star which is located just beyond the critical
point of instability. In this case, we have $T/W \approx 0.0088$ 
and $q \approx 0.96$, values 
which are nearly equal to those of the critical configuration (see
Table \ref{t-1}). 
A second simulation was performed for a more compact star 
with $R \approx 570M$, but our results were
essentially unchanged. 
To accelerate the collapse, we depleted the pressure by 1\%
initially. During the evolution, we adopted 
a $\Gamma$-law equations of state according to $P=(\Gamma-1)\rho\varepsilon$ 
where $\varepsilon$ denotes the specific internal energy of the fluid, 
thereby treating the gas as an ideal, adiabatic fluid in which
cooling can be ignored on a dynamical timescale
(Linke et al. 2001). 

Since the equatorial radius decreases by a factor of $\sim 1000$ 
(from $\sim 600M$ to $<M$), using a fixed uniform grid with sufficient
resolution for all epochs would be computationally prohibitive.
Instead, we adopted a uniform grid with 
decreasing grid spacing and increasing grid number as the collapse proceeded 
in order to guarantee adequate resolution up to the 
formation of an apparent horizon.
For the early stages, where the radius at the equatorial surface  
$R_m$ exceeds $150M$, we used $N=300$ grid points in the 
$x$ and $z$ directions. Here, we identify  $R_m$ as the radius
at which $\rho=10^{-6}\rho_{\rm max}$, where $\rho_{\rm max}$ is the maximum
interior density. Initially, the grid covers the equator 
with about 200 grid points. 
The outer boundaries
along the $x$ and $z$ axes are located at $L \approx 930M$. 

Since $T/W$ is small and $\Gamma = 4/3$, 
the collapse proceeds in a
homologous manner in the central regions during the early stages 
(Shapiro \& Teukolsky 1979; Goldreich \& Weber 1980; Saijo et al. 2002). 
Taking advantage of this behavior, we rezoned by moving the outer boundary
inward and decreasing the grid spacing, keeping $N(=300)$ fixed. 
All quantities in the new grid are calculated using  cubic interpolation. 
We discarded the outermost 
computational domain, but the discarded baryon rest-mass is very small 
(less than $10^{-4}$ of the total) 
when $R_m \agt 150M$. We repeated this procedure twice until homology  
breaks down at $R_m \alt 150M$.
After this stage, the collapse timescale in the central region
is much shorter than in the envelope. 
Consequently, we increased the grid number $N$ and decreased $\Delta$, 
monitoring the lapse function at the center $\alpha_0$.
Specifically, we set $N$ and $L$ as follows:
for $\alpha_0 \geq 0.90$, we set $N=600$ and $L \approx 233M$;
for $0.70 \leq \alpha_0 \leq 0.90$, we set $N=900$ and $L \approx 155M$;
and for $0.30 \leq \alpha_0 \leq 0.70$, we set $N=1200$ and $L \approx 116M$.
With this treatment, the discarded fraction of the baryon rest-mass is 
only $\sim 0.15 \%$ down to $\alpha_0=0.3$. 

For $\alpha_0  \alt 0.3$, 
the central density profile becomes very steep. To see its dependence on 
grid resolution during black hole formation, 
we carried out simulations using different combinations of 
$N$ and $L$ to refine the grid, with the restriction that $N \leq 2400$. 
The set-up of each simulation was as follows: 
(A) for $\alpha_0 \leq 0.30$, we took $N=1200$ and $L \approx 116M$,
with no regridding at $\alpha_0=0.3$; 
(B) for $\alpha_0 \leq 0.30$, we took $N=1200$ and $L \approx 58M$,
regridding once at $\alpha_0=0.3$; 
(C) for $0.05 \leq \alpha_0 \leq 0.30$, we took $N=1200$ and $L \approx 58M$ and
for $\alpha_0 \leq 0.05$, we took $N=1200$ and $L \approx 29M$;  
(D) for $0.05 \leq \alpha_0 \leq 0.30$, we took $N=1200$ and $L \approx 58M$ and
for $\alpha_0 \leq 0.05$, we took $N=2400$ and $L \approx 58M$.
For cases (C) and (D), we carried out regridding 
at $\alpha_0=0.3$ and 0.05. The minimum grid spacings for
cases (A)--(D) were $0.097M$, $0.048M$, $0.024M$ and $0.024M$, respectively. 
The results for case (D) are 
the most reliable, but the results 
for the four cases do not differ significantly.

For cases (B)--(D), 
the outer boundaries reside deep inside the stellar surface. 
Hence, the fluid at large 
equatorial radius is discarded.
When we allow $L \approx 58M~(29M)$, the loss of  
total baryon rest-mass becomes $\sim 2 \%~(4 \%)$.
Since the outer envelope 
has larger specific angular momentum, setting $L\approx 58M~(29M)$
implies that $\sim 8\%~(18 \%)$ of the total angular momentum is lost. 
As shown below, however, discarding some mass  
in the outer region is a tiny effect on the formation and evolution of a SMBH
in the central region. 

\section{Numerical Results}

In Fig. 1, we display snapshots of density contours 
and velocity vectors in the $x$ - $z$ plane
at selected times. Here, the velocity field is
defined as $u^i/u^t$ where $u^{\mu}$ is the four velocity.
The collapse proceeds nearly homologously
in the central region during the early stages.
However, for $R_m \alt 150M$, 
the effects of rotation and general relativity modify this property, 
when the collapse near the central region 
accelerates significantly.  

In Fig. 2, we show the time evolution of the
central conformal factor ($\psi_0\equiv [{\rm det}(\gamma_{ij})]^{1/12}$
at $r=0$ 
where $\gamma_{ij}$ denotes the three-dimensional spatial metric)
and $\alpha_0$ for cases (A)--(D). 
We find that the collapse proceeds in a
runaway manner in the final stages, although the time development
depends somewhat on our adopted choice of time slicing. 

Since $\psi^2\cdot\Delta$  measures the proper physical length of the 
grid spacing, 
maintaining the resolution requires changing $\Delta$ as
$\Delta \propto \psi^{-2}$ (i.e, $N \propto \psi^2$). 
However, $\psi$ diverges very sharply at late times, 
so increasing $N$ by a factor of a few does not 
improve the resolution much. Accordingly, we
terminated the simulation when the conservation of $M$ 
was violated by $\sim 10\%$ at $t/M \approx 30636$.

There are two reasons for this runaway behavior at the center.
One is that the equation of state is very soft, which produces
stars with very centrally condensed
structures. The collapse timescale in the
central region is then much shorter than that in the outer
region. The other reason is 
our choice of a coordinate (shift) condition, for which 
the resolution near the center deteriorates ( ``grid sucking''; 
see Shibata 1999b). 
Integrating to a final equilibrium state evidently requires
different time and/or spatial gauge conditions.

We find an apparent horizon forms for $t/M \agt 30630$.
In Fig. 3, we show the mass of the apparent horizon 
as a function of time.
The mass is defined as
$M_{\rm AH} \equiv \sqrt{A/16\pi}$ where 
$A$ denotes the area of the apparent horizon 
(e.g., Cook \& York, 1990). In addition to $M_{\rm AH}$, 
we plot the total baryon rest-mass inside the apparent 
horizon $M_{*{\rm AH}}$ for case (D) (dotted curve),
which agrees approximately with $M_{\rm AH}$. 
Figure 3 indicates that at the end of the simulation 
about 60 \% of the total rest-mass already has been 
swallowed into the black hole. Clearly, {\it collapse of a rapidly and 
uniformly rotating SMS leads to formation of a SMBH.} 
However, our final snapshot is
not the final state, because $M_{\rm AH}$ is still increasing. 

It is possible to estimate what the final mass and spin of the
black hole will be once it settles into equilibrium.
Define the specific angular momentum according to 
$j\equiv h u_{\varphi}$, where $h(=1+\varepsilon+P/\rho)$ is the specific
enthalpy. In an axisymmetric system,
the integrated baryon rest-mass of all fluid elements with $j$ less than 
a given value $j_0$, $m_*(j_0)$ (the angular momentum spectrum), is 
conserved in the absence of 
viscosity (Stark \& Piran 1987): 
\beq
m_*(j_0)=\int_{j<j_0} \rho \alpha u^t \psi^6 d^3x. 
\eeq
Consider the innermost stable circular orbit (ISCO)
around the growing black hole at the center.  
If $j$ of a fluid element is less than
the value at the ISCO $(j_{\rm ISCO})$,
the element will fall into the black hole eventually.
Now the possibility exists that some fluid can be captured 
even for $j > j_{\rm ISCO}$, if it is in a noncircular orbit. 
Ignoring these trajectories  yields the minimum amount of mass 
that will fall into the black hole at each moment. 
The value of $j_{\rm ISCO}$ changes 
as the black hole grows. 
If $j_{\rm ISCO}$ increases,  additional mass  
will fall into the black hole. However,
if $j_{\rm ISCO}$ decreases,  ambient fluid
will no longer be captured. This expectation 
implies that when $j_{\rm ISCO}$ reaches a maximum value, 
the dynamical growth of the black hole will terminate. 

To analyze the  growth of the black hole mass, we generate Figs. 4 and 5. 
In Fig. 4, we show the angular momentum spectrum 
at $t=0$. To verify that the spectrum is well preserved during 
the simulation, we also plot the spectrum at the time when the apparent horizon 
is first formed. In Fig. 5 (a), we plot 
$q_*\equiv J(j)/m_*(j)^2$ as a function of $m_*(j)/M_*$. Here, 
$J(j)$ denotes the total angular momentum 
with the specific angular momentum $j$ less than a 
given value $j_0$ and is defined according to 
\beqn
J(j_0)=\int_{j<j_0} \rho \alpha u^t \psi^6 h u_{\varphi}d^3x. 
\eeqn
Now, $J(j_{\rm ISCO})/m_*(j_{\rm ISCO})^2$ and 
$m_*(j_{\rm ISCO})$ may be approximately 
regarded as the instantaneous spin parameter and mass of a black hole.
Therefore, the dotted curve in 
Fig. 5(a), derivable from the initial stellar profile,
may be interpreted as an approximate evolutionary track 
for the angular momentum parameter of the growing black hole. 
The numerical results for various resolutions indicate that 
this is close to the actual track followed by the hole. Thus, 
we can assume that 
our assumptions made in this analysis are adequate.
The solid curve in Fig. 5(a) shows 
that with the increase of the black hole mass, 
the spin parameter also increases. 

If we assume that $m_*(j)$ and $q_*$ are the  
mass and spin parameter of 
the black hole and that the spacetime 
can be approximated instantaneously by a Kerr metric, 
we can compute $j_{\rm ISCO}$  
(see, e.g., chapter 12 of Shapiro \& Teukolsky, 1983). 
In Fig. 5(b), we show $j_{\rm ISCO}[m_*(j),q_*(j)]$ 
as a function of $m_*(j)$. For $m_*(j) < 0.9M_*$, 
$j_{\rm ISCO}$ is an increasing function of the mass.
This implies that the mass of the black hole should increase to
$\sim 0.9M_*$. However, 
at $m_*(j)/M_* \approx 0.9$, $j_{\rm ISCO}$ reaches a maximum, 
as $j_{\rm ISCO}/m_*(j)$ steeply decreases 
above this mass fraction.
Thus, once the black hole reaches this point, it will stop
growing dynamically.  Figure 5(a) shows that at this stage, $q_* \sim 0.75$. 
Therefore, at the end of this collapse, 
(1) about 90\% of the total mass will form a SMBH 
and (2) the spin parameter of the Kerr hole 
at the end of the collapse will be $\sim 0.75$. 

\section{Summary} 

We performed a fully relativistic numerical simulation
in axisymmetry of the collapse of a uniformly rotating, marginally
unstable SMS. 
Our simulation terminates when
roughly 60\% of the mass has been swallowed by the SMBH. 
We estimate that about 90\% of the total mass of
the system will be consumed by the end of the collapse.
The spin parameter $q$ of
the final Kerr SMBH is likely to be $\sim 0.75$ at the end of the 
dynamical collapse phase. 
Most of the remaining gas will reside in an ambient disk about the
central hole.

To follow black hole growth to a final equilibrium state will 
require different coordinate gauge choices for the lapse and shift 
functions. It may also
require the use of an ``horizon excision'' boundary condition
(Unruh, unpublished; Thornburg 1987; Seidel \& Suen 1992).
To simultaneously follow the extended disk may
necessitate employing a nested grid or adaptive mesh refinement. 
These issues are ripe for future exploration.

\acknowledgments

We are grateful to T. Baumgarte, Y. Eriguchi, 
H. Shinkai, H. Susa, M. Umemura, and K. Uryu for discussion. 
Numerical computation was performed on FACOM VPP5000
in the data processing center of National Astronomical
Observatory of Japan. 
This work was in part supported by a Japanese
Monbu-Kagaku-sho Grant (No. 13740143) and by NSF Grant
PHY-0090310 and NASA Grants NAG5-8418 and NAG5-10781 at the
University of Illinois at Urbana-Champaign.


\begin{table}
\begin{center}
\caption{Parameters of the initial SMS.\label{t-1}}
\begin{tabular}{cccccccc}
\tableline\tableline
 & $R/M$ & $T/W$ & $q$ & $M$ & $M_*$ & $\Omega$ & $\rho_c$ \\
\tableline
 initial data & 622 & 0.0088 & 0.96 & 4.566 & 4.566 & 1.40e-5 &
7.84e-9 \\
\tableline
\end{tabular}
\tablecomments{\rm From left, the radius at the equator,
ratio of the kinetic energy to the gravitational binding
energy, gravitational mass, baryon rest-mass, angular
velocity and central density. All the quantities are shown
in units of $c=G=K=1$.}
\end{center}
\end{table}


\begin{figure}
\begin{center}
\epsfxsize=1.68in
\leavevmode
\epsffile{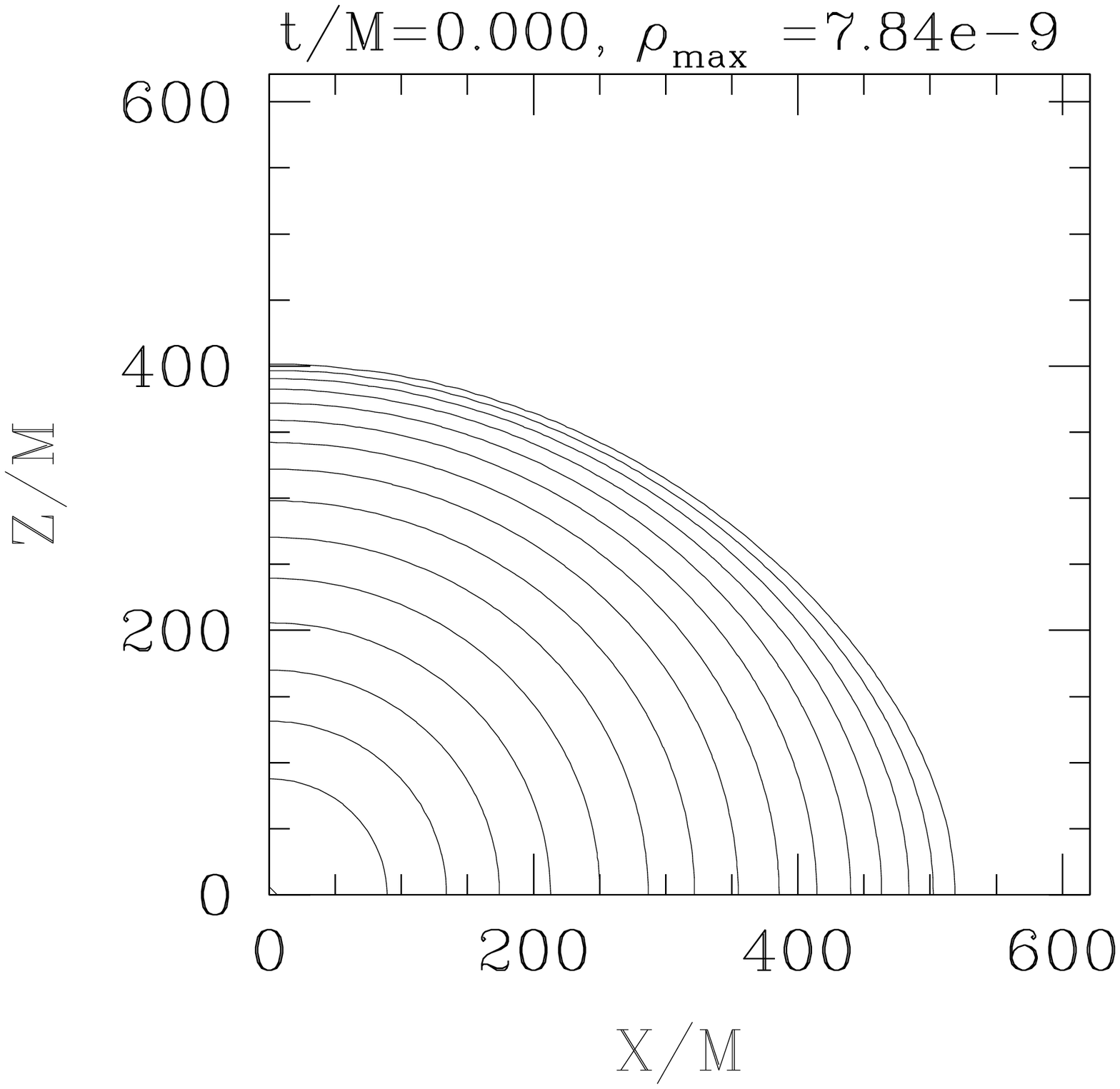}
\epsfxsize=1.68in
\leavevmode
\epsffile{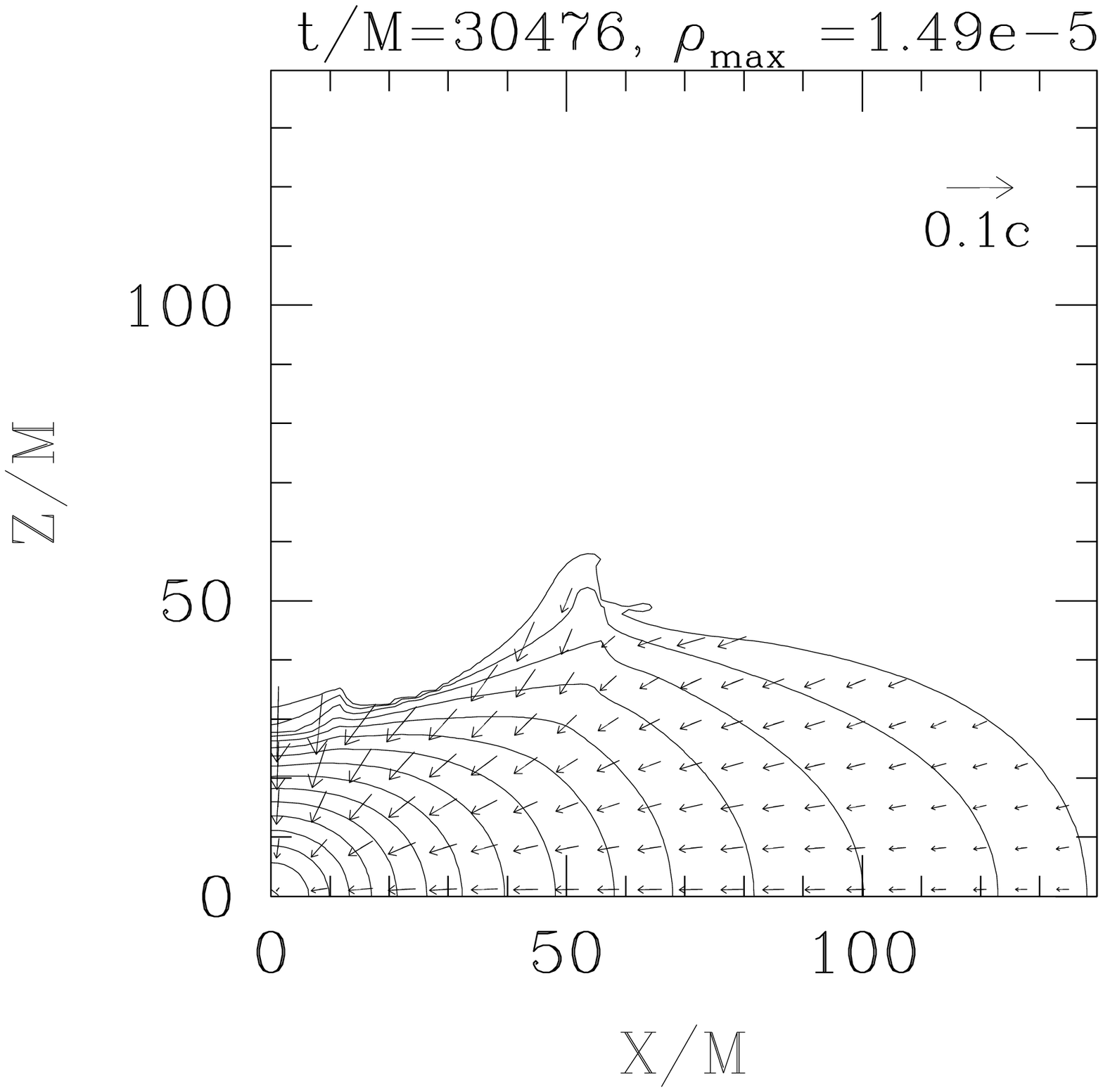}
\epsfxsize=1.68in
\leavevmode
\epsffile{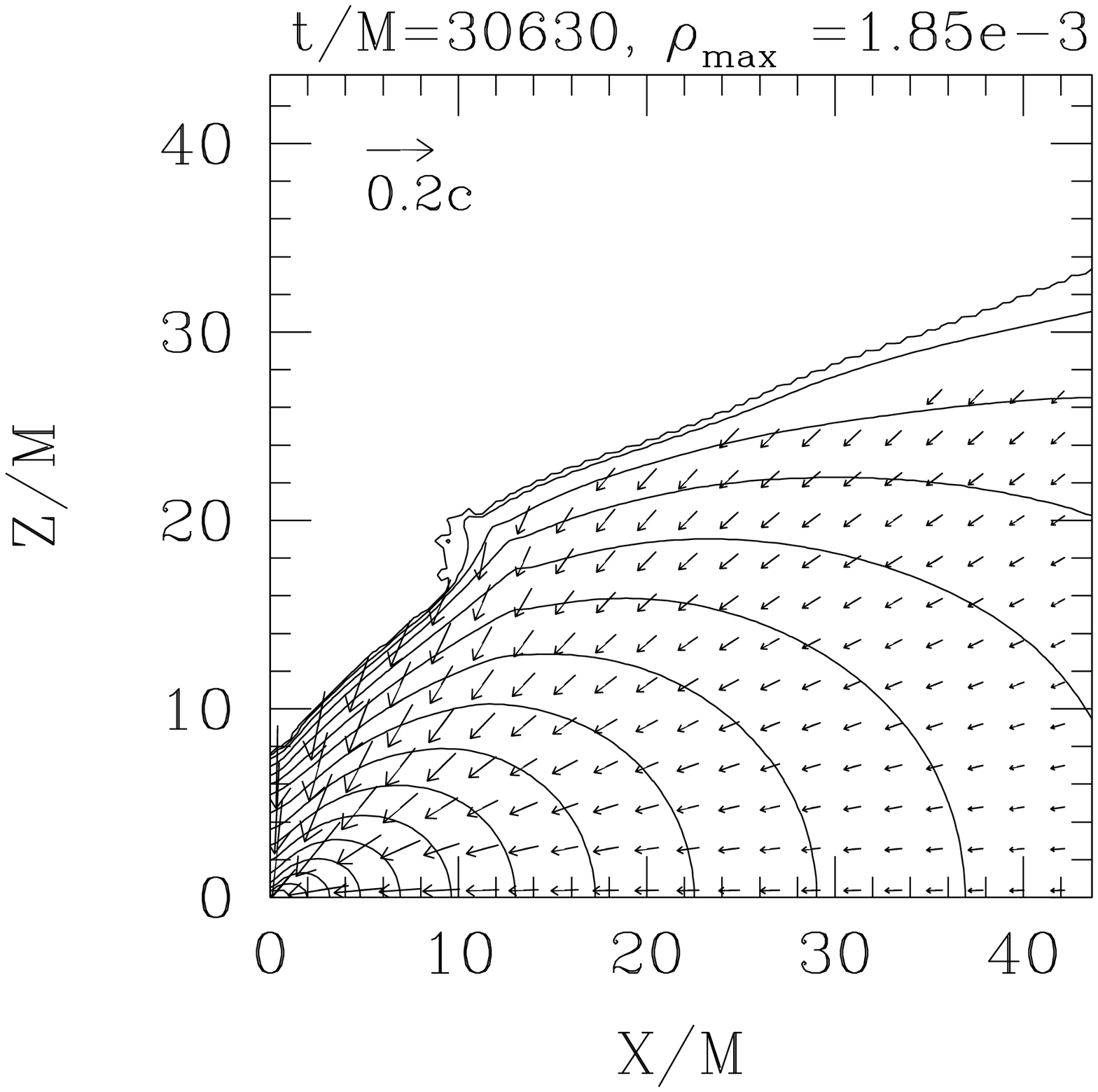}
\epsfxsize=1.68in
\leavevmode
\epsffile{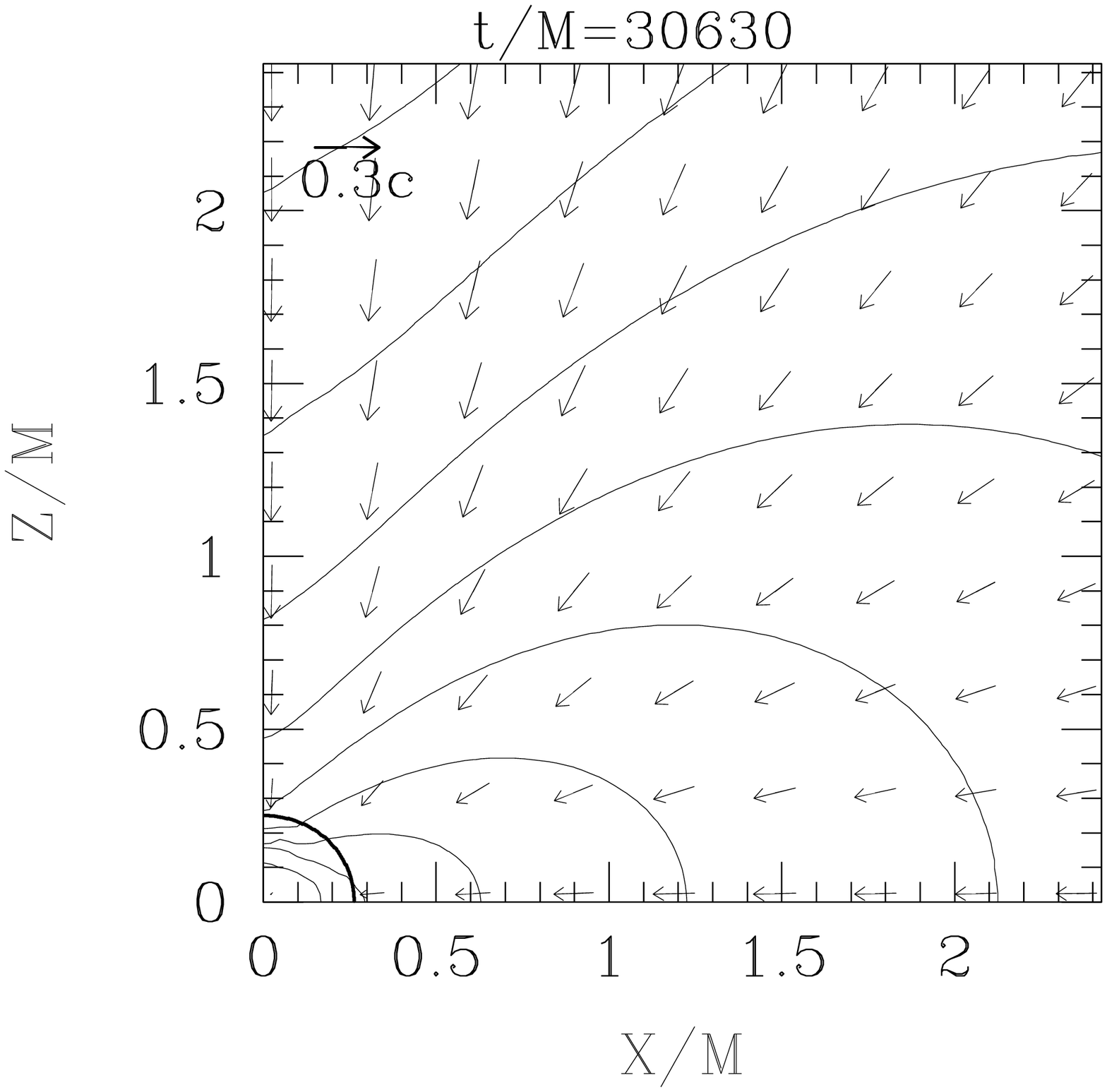}
\end{center}
\caption{Snapshots of density contours and velocity vectors in the
$x$-$z$ plane at selected times for case (D).
The contours are drawn for $\rho/\rho_{\rm max}
=10^{-0.4j}~(j=0 \sim 15)$, 
where $\rho_{\rm max}$ denotes the maximum density at each time.
The fourth figure is a blow-up of the third one in 
the central region: 
The thick solid curve at $r \approx 0.3M$ shows the location of 
the apparent horizon. 
\label{fig1}}
\end{figure}

\begin{figure}
\begin{center}
\epsfxsize=2.5in
\leavevmode
\epsffile{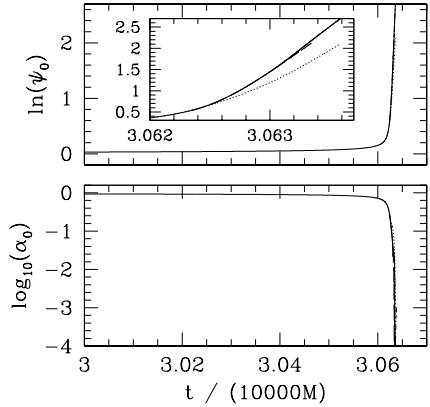}
\end{center}
\caption{The central value of the conformal factor $\psi_0$
and lapse function $\alpha_0$
as a function of time. The solid, dotted, dashed, and
dotted-dashed curves denote the results for cases (D), (A), (B) and (C).
The results for (C) and (D) are not
distinguishable. \label{fig2}}
\end{figure}

\begin{figure}
\begin{center}
\epsfxsize=2.2in
\leavevmode
\epsffile{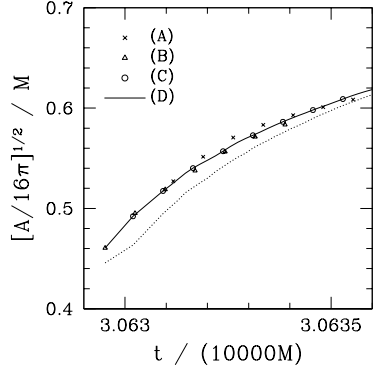}
\end{center}
\caption{Evolution of the  mass of apparent horizon (solid curve) and
baryon rest-mass inside the apparent horizon ($M_{*{\rm AH}}$, dotted curve)
as a function of time for case (D).
Mass and time are shown in units of the total 
gravitational mass $M$. For comparison,
we also plot the mass of apparent horizon as a function of time
for cases (A) (crosses), (B) (triangles) and (C) (circles).
The results for cases (B), (C) and (D) essentially agree. \label{fig3}}
\end{figure}

\begin{figure}
\begin{center}
\epsfxsize=2.5in
\leavevmode
\epsffile{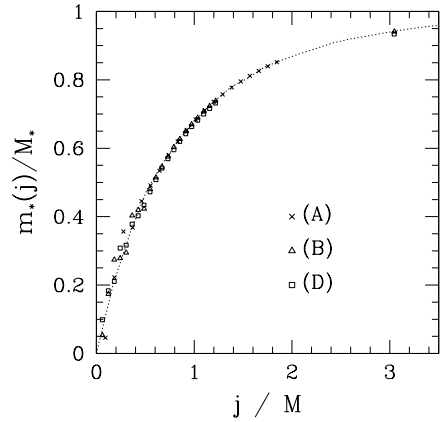}
\end{center}
\caption{The specific angular momentum spectrum at
$t\approx 0$ (dotted curve)
and at the first formation of an apparent horizon
at $t\approx 30630M$ for cases (D) (squares), (A) (crosses) and
(B) (triangles). \label{fig4}}
\end{figure}

\begin{figure}
\begin{center}
\epsfxsize=3.in
\leavevmode
\epsffile{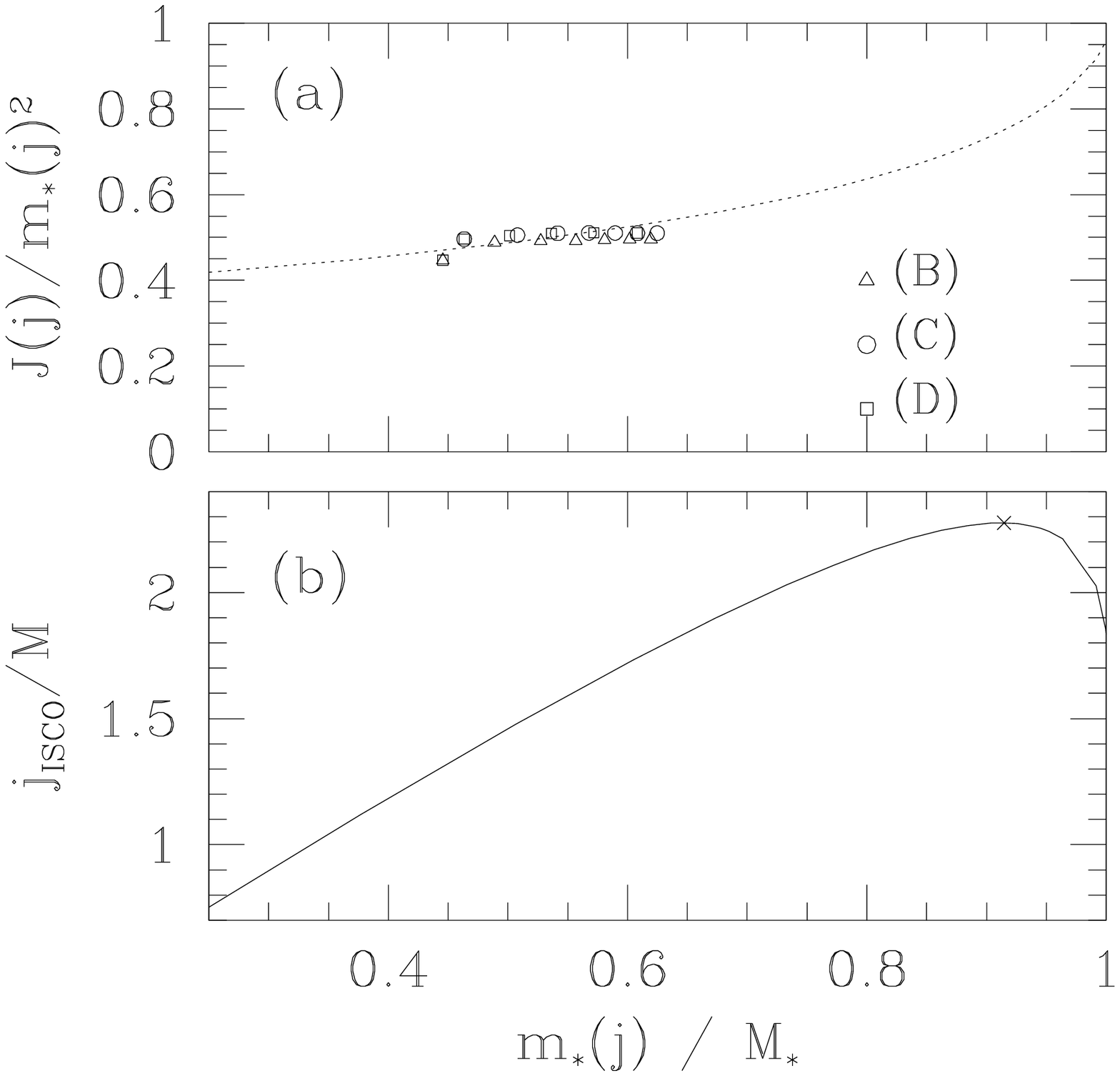}
\end{center}
\caption{(a): $J(j)/m_*(j)^2$ as a function of $m_*(j)$ (dotted curve). 
The squares, circles and triangles show $J_{\rm AH}/M_{*{\rm AH}}^2$ and
$M_{*{\rm AH}}$ at select times 
for cases (D), (C) and (B), respectively. 
Here, $J_{\rm AH}$ and
$M_{*{\rm AH}}$ are the angular momentum and baryon rest-mass
inside the apparent horizon.
(b): $j_{\rm ISCO}/M$ as a function of $m_*(j)/M_*$.
The cross marks the maximum. 
\label{fig5}}
\end{figure}


\begin{thebibliography}{}
\bibitem{alc01} Alcubierre, M. et al., 2001, Int. J. Mod. Phys. D 
10, 273.
\bibitem{bal02} Balberg, S and Shapiro, S. L., 2002, 
Phys. Rev. Lett., in press (astro-ph/0111176). 
\bibitem{bau99} Baumgarte, T. W. \& Shapiro, S. L., 1999, ApJ, 
526, 941.
\bibitem{beg78} Begelman, M. C. \& Rees, 1978, M.J., MNRAS, 
185, 847.
\bibitem{coo90} Cook, G. B. \& York, J. W., 1990, Phys. Rev. D 41, 1077. 
\bibitem{gol80} Goldreich, P. \& Weber, S. V., 1980, ApJ, 238, 991.
\bibitem{lin01} Linke, F., Font, J. A., Janka, H.-T., M\"uller, E. \&
Papadopoulas, P., 2001, A\&A, 376, 568.
\bibitem{loe94} Loeb, A. \& Rasio, F. A., 1994, ApJ. 432, 52.
\bibitem{new01} New, K. C. B. \& Shapiro, S. L., 2001, ApJ. 548, 439.
\bibitem{qui90} Quinlan, G. D. \& Shapiro, S. L., 1990, ApJ, 356, 483.
\bibitem{ree84} Rees, M. J., 1984, ARA \& A, 22, 471. 
\bibitem{ree98} Rees, M. J., 1998, 
in {Black holes and relativistic stars}, ed.
R. M. Wald (Chicago University Press),  79.
\bibitem{ree01} Rees, M. J., 2001, 
in {Black holes in Binaries and Galactic Nuclei}, ed.
L. Kaper, E. P. J. van den Heurel, \& P. A. Woudt (New York: 
Springer-Verlap), 351.
\bibitem{sai01} Saijo, M, Baumgarte, T. W., Shapiro, S. L. \&
Shibata, M, 2002, ApJ, in press. 
\bibitem{sch01} Schutz, B. F., 2001, gr-qc/0111095. 
\bibitem{sei92} Seidel, E. \& Suen, W., 1992, Phys. Rev. Lett. 69, 1845. 
\bibitem{sha83} Shapiro, S. L. \& Teukolsky, S. A., 1983, 
{Black  Holes, White Dwarfs, and Neutron Stars} 
(Wi1ey interscience, New York).
\bibitem{sha85} Shapiro, S. L. \& Teukolsky, S. A. 1985,
ApJ, 292, L91.
\bibitem{shi99a} Shibata, M., 1999a, Prog. Theor. Phys. 101, 1199.
\bibitem{shi99b} Shibata, M., 1999b, Phys. Rev. D, 60, 104052.
\bibitem{shi00} Shibata, M., 2000, Prog. Theor. Phys., 104, 325.
\bibitem{sta87} Stark, R. F. \& Piran, T., 1987, Comp. Phys. Rep., 5, 221. 
\bibitem{tho87}  Thornburg, J., 1987, Class. Quantum Grav. 4, 1119. 
\bibitem{tho95} Thorne, K. S., 1995, in {Proceeding of Snowmass 95
 Summer Study on Particle and Nuclear Astrophysics and Cosmology},
eds. E. W. Kolb and R.  Peccei (World Scientic, Singapore), 398.
\bibitem{ume01} Umemura, M., 2001, ApJ, 560, L29.   
\bibitem{zel71} Zel'dovich, Ya. B. \& Novikov, I. D., 1971, 
{Relativistic Astrophysics Vol. 1} (University of Chicago Press).
\end{thebibliography}
\end{document}